\documentstyle[preprint,eqsecnum,aps]{revtex}

\begin{document}
\draft
\preprint{TD Bank and UW}
\title{Pricing defaultable debt: some exact results}
\author{D. F. Wang$^{1,2}$}
\address{$^1$Department of Statistics and Actuarial Science\\
University of Waterloo, 
Waterloo, ONT N2L 3E5 Canada
\\
$^2$Toronto Dominion Bank}
\date{May, 1998}
\maketitle
\begin{abstract}
In this letter, I consider the issue of pricing risky debt
by following Merton's approach. I generalize Merton's results to the case
where the interest rate is modeled by the CIR term structure.
Exact closed forms are provided for the risky debt's price.
\end{abstract}

%%%%%%%%%%%%%%%%%%%%%%%%%%%%%%%%%%%%%%%%%%%%%%%%%%%%%%%%%%%%%%%%%%%%%%%
%\narrowtext

%\setcounter{page}{27}

There are considerable recent interests in pricing derivatives
with credit risk, among theoretical researchers and practitioners.
In general, pricing instruments with credit risk follows two
approaches. The first one, pioneered by Merton\cite{merton1,shimko,long,longstaff,zhou}, 
is to assume a 
stochastic process for the value of the firm, and then handle the 
risky debt as an option. In particular, a very interesting
jump-diffusion approach to credit risk was proposed recently\cite{zhou}
within this framework.
The second approach is to assume stochastic processes
for the credit quality of each bond, and the recovery rate in the event
of default\cite{jarrow}.
In this letter, I follow the approach of Merton's. The problem of pricing risky
debt is studied when the interest rate is modeled by the Cox-Ingersoll-Ross term
structure. Closed form results are provided for the prices of the risky debt.
In Merton's original work, he assumed that the interest rate is constant
and that the default event of the debt can only occur at the time of 
maturity. He then gave the closed form of the risk debt's price.
As stated explicitly in his work\cite{merton1}, one can generalize the results
to the case when interest rate is stochastic. This can easily 
be achieved by using Merton's work that generalized Black-Scholes formular
to the option pricing with stochastic interest rate. 

However, one 
has to make the assumption that the bond process of the stochastic 
interest rate will have to have a non-stochastic volatility which is 
allowed to be deterministic time-dependent\cite{merton1}.
Among well-known interest rate models, such as Vasicek model\cite{vasicek} and 
Cox-Ingersoll-Ross model\cite{cox2}, the Vasicek model will give rise to
a non-stochastic volatility for the corresponding bond process, while
the Cox-Ingersoll-Ross will give a stochastic bond volatility.
Henceforth, Shimko etc \cite{shimko} applied Merton's results to the case of 
stochastic interest rate described by the Vasicek model. Closed forms
of the price of the risky debt can be given exactly. Longstaff
and Schwartz also considered the problem where default event can occur
before maturity and a Vasicek interest rate model was used\cite{long}.
It has remained open 
for a long time whether one can derive  a similar closed form result 
for the risky debt's price when the interest rate is CIR term structure.
The results of this letter partly fill this open gap that remains 
for a long time. 

Assume a probability space denoted by
$(\Omega, P, \{F_t\}, F)$, with the filtration $\{F_t\}$.
Consider the value of the firm that is described by the following process
\begin{equation}
{dV\over V} =\mu dt + \sigma d Z_1, 
\end{equation}
where $Z_1$ is a Brownian motion in the probability space.
The interest rate process is assumed to be the one given by Cox-Ingersoll-Ross\cite{cox2}:
\begin{equation}
dr=(a -\beta r) + \eta  dZ_2,
\end{equation}
where $\eta =\sigma_r   r^{1/2}$ with $\sigma_r$ as a constant. 
For this term structure,
they have computed the riskless bond price explicitly.
The co-quadratic variational process is $[Z_1,Z_2]=\rho t$. 

The firm issues equity and debt. The assumptions in Merton's paper
are also made here\cite{merton1}. 
The total value of the firm is 
the sum of equity and debt. The PDE satisfied by the equity 
is given by 
\begin{equation}
H_{\tau}={\sigma^2\over 2} V^2 H_{VV} +\rho \eta \sigma VH_{Vr}
+{\eta^2\over 2}H_{rr} +r VH_V +(\alpha -\beta r) H_r -r H
\end{equation}
where $H=H(V, r, T-t)$ and $\tau=T-t$ is time to the maturity, and 
$\alpha$ is sum of $a$ plus the constant representing 
the market price of the interest rate risk.
At $\tau=0$, the equity should satisfy the boundary condition
that $H=max(0, V(T)-B)$, where $B$ is the face value of the debt issued 
by the firm maturing  at time $T$. The risky debt price is therefore
given by $Y=V(t)-H(V,r,T-t)$. For simplicity, we assume, as Merton did, 
that event of default of the risky debt can only occur at the time of 
maturity.   

Following the standard risk neutral approach, we write the equity 
price as below:
\begin{equation}
H=E^Q(e^{-\int_t^T r(s) ds} max(0,V(T)-B)|F_t), 
\end{equation}
where the expectation $E^Q$ means that in the risk-neutral-adjusted world.
In this risk-neutral world, the firm value and the interest rate will follow
the stochastic differential equations as 
\begin{eqnarray}
&&d lnV=(r-{1\over 2} \sigma^2) dt +\sigma d\hat Z_1\nonumber\\
&& d r =(\alpha-\beta r )dt +\eta d\hat Z_2.
\end{eqnarray}
Here, both $\hat Z_1$ and $\hat Z_2$ are Wiener processes in the risk-neutral
world, and the co-quadratic process is $[{\hat Z_1}, {\hat Z_2}]= \rho t $.

In the special case where $Z_1$ and $Z_2$ are independent of each other,
$\hat Z_1$ and $\hat Z_2$ are independent of each other, i.e. $\rho=0$.
In this case, we may find explicit closed form for the equity price.
Given a sample  path of the interest rate $\{r(s)\}$ for $s\in [t,T]$, we know
that $m=\int_t^T r(s) ds $ is given. Hence, conditional on that 
the sample path of the interest rate $r$ is given, we have
\begin{equation}
E^Q(e^{-\int_t^T r(s) ds} (V(T)-B)^+|F_t,\{r(s)\})
=e^{-m}E^Q( (V(T)-B)^+ |F_t, \{r(s)\}). 
\end{equation}
This conditional expectation can be carried out with the standard
method. It is found to be 
\begin{equation}
V(t)N(d_1)-Be^{-m} N(d_2)
\end{equation}
where $d_{1,2}={ ln(V(t)/B)+ (m\pm \sigma^2 (T-t)/2 ) \over \sigma 
\sqrt{T-t} }$. Therefore, the price of the equity will be expressed 
in terms of the probability density of $m$. Denote $g(m)$ the density for  
the random variable $\int_t^Tr(s)ds$ to be in the region $[m, m+dm]$.
The equity price will thus take the form 
\begin{equation}
H(V(t), r(t), T-t)=\int_0^{\infty} (V(t)N(d_1) -B e^{-m} N(d_2)) g(m)dm.
\end{equation}
The remaining task is to find the closed form of the density distribution
function $g(m)$. 

For the Cox-Ingersoll-Ross term structure, the density function $g(m)$ can be 
found easily. Consider the moment generating function   
\begin{equation}
I(x)=E^Q(e^{-x \int_t^T r(s) ds} |F_t)=\int_0^\infty e^{-mx} g(m)dm, 
\end{equation}
where $x$ is any non-zero numbers. Using the bond price of Cox-Ingersoll-Ross,
we find the moment generating function. 
For CIR term structure, stochastic differential equation governing 
the short rate is $dr = (\alpha-\beta r )dt +\sigma_r \sqrt{r} d\hat Z_2$,
will remain unchanged under the scaling transformation
\begin{eqnarray}
&&r\rightarrow x r\nonumber\\
&&\beta \rightarrow \beta\nonumber\\
&&\alpha\rightarrow x  \alpha\nonumber\\
&&\sigma_r\rightarrow x^{1/2} \sigma_r,
\end{eqnarray} 
where $x$ is any positive real number.
Denote $D(r(t); \alpha, \beta, \sigma_r, T-t)=E^Q(e^{-\int_t^Tr(s)ds}|F_t)$  
the riskless zero-coupon bond  
price at time $t$ whose payoff at maturity $T$ is one. The exact closed
form of this zero-coupon bond price was provided explicitly\cite{cox2}.
We obtain 
\begin{equation}
I(x) =D(xr(t), x\alpha, \beta, x^{1/2} \sigma_r, T-t)=
\int_0^\infty g(m) e^{-xm}dm.
\end{equation}
Doing inverse transformation, we will be able to find the density function  
$g(m)$. Substituting this density function into our previous pricing formular,
we therefore obtain the closed form of the price of the equity.
The price of the risky debt $Y$ is equal to $V(t)$ minus 
the equity value, i.e. $Y(t)=V(t)-H$. In terms of the density distribution  
$g(m)$, we have 
\begin{equation}
Y(t)= \int_0^\infty (V(t)-[V(t)N(d_1)-Be^{-m}N(d_2)]) g(m) dm.
\end{equation}
The credit spread 
is given by $r_D=-{1\over T} ln (Y/B) -r(t)$. The credit spread is an 
increasing function of the face value $B$, but a decreasing function
of the firm's asset value $V(t)$. 
 
We can generalize these closed form results to the following cases. 
First, one can consider the case where 
the volatility of the firm's value is a time dependent deterministic 
function $\sigma(s)$. We simply replace $\sigma^2 (T-t)$ in our final 
formular by the integral $\int_t^T \sigma^2(s) ds$. This will give 
us the closed form for the price of the risky debt.

Second, we can generalize our results 
when the firm value $V$ is modeled by the CEV processes
$dS =\mu S dt + \sigma S^{\alpha/2} dZ_1$. Cox derives a closed form 
solution for the European option\cite{cox3}. When the interest rate is modeled 
with CIR term structure, we can derive the closed form
solution for the risky debt when the Wiener process $Z_1$ is independent of 
the interest rate's Wiener process $Z_2$. 

Third generalization is 
when the firm value has a stochastic volatility. We may assume that the 
firm value is modeled as 
\begin{eqnarray}
&&{dS\over S} = \mu dt +\sigma dZ_1\nonumber\\
&&d \sigma^2 =(\eta-\beta \sigma^2) dt +\sigma \nu dZ_2\nonumber\\
\end{eqnarray}
where the second equation governs the behavior of the 
volatility of the firm's value. When the term structure is modeled 
by the CIR model: $dr =(a-br)dt + \sigma_r \sqrt{r} dZ_3$,
we may also find the closed form of the risky debt in the special case where
$Z_1, Z_2$ and $Z_3$ are independent  of each others. 
In this case, we readily find the moment generating function  
for the interest rate part, as well as the moment generating function for 
the $\sigma^2$. In fact, given the sample paths of the interest rate,
and given the sample path of the variance of the firm's value, one can price
the value of the equity with standard method. 
Averaging this over the distributions of 
the interest rate sample path and 
over the $\sigma^2$ sample path by using the corresponding 
moment generating functions, we can derive the closed form
for the equity price. Once this is done, we will easily find the 
price of the risky debt and the corresponding credit spread.

In summary, we have made an attempt to price risky debt by following Merton's 
approach. An effort has been taken by me 
to go beyond Merton's closed form results
which require that 
the riskless bond's volatility is a non-stochastic function of 
time. Our results partly fill the open gap that remains for a long time.
Our results can easily apply to any term structure of riskless interest rate
for which a closed form of zero-coupon bond price has been found.
It still remains to find a similar closed form solutions of risky debt
when the Wiener process of the firm value has non-zero co-quadratic variational
process with the Wiener process of the interest rate. 
A detailed account based on this letter will be 
reported in further works.

Email address:d6wang@barrow.uwaterloo.ca.
I am indebted to Professors P. Boyle and D. McLeish for
the finance theories I learned from them. Conversations with
Prof. K. S. Tan of UW,
Dr. Hou-Ben Huang and Dr. Z. Jiang of TD Securities, Dr. Bart Sisk and Dr. A. Benn of TD Bank,
Professor K. Wang of University of Chicago Graduate School of Business and University of McGill, 
Dr. Daiwai Li
and Dr. Craig Liu of Royal Bank, Dr. ChongHui Liu  and Dr. J. Faridani
of Scotia Bank, are gratefully knowledged. I also wish to thank Dr. Rama Cont
for informative communication. The opinions of this article are those of the
author's, and they do not necessarily reflect the institutions the author is affiliated
with. Any errors of this article are mine.

\end{document}